\documentclass[12pt]{article}
\usepackage[utf8]{inputenc}
\usepackage{amscd,amsfonts,amstext,amsmath,amssymb,amsthm,bm}
\usepackage[numbers]{natbib}
\usepackage{mathtools}
\usepackage{subcaption}
\usepackage[margin=2cm]{caption}
\usepackage{url}
  \newenvironment{dedication}
        {\vspace*{3ex}\begin{quotation}\begin{center}\begin{em}}
        {\par\end{em}\end{center}\end{quotation}}

\DeclareMathOperator*{\argmin}{argmin}

\newcommand{\esp}{\mathbb{E}}
\newcommand{\var}{\mathbb{V}}

\newcommand{\lossf}{\mathcal{L}}
\newcommand{\riskf}{\mathcal{R}}
\newcommand{\sigmaf}{\mathcal{F}}
\newcommand{\sigmag}{\mathcal{G}}
\newcommand{\dkappa}{\dot{\kappa}}
\newcommand{\realn}{\mathbb{R}}
\newcommand{\diag}{\mbox{diag}}
\newcommand{\nat}{\mathbb{N}}
\newcommand{\ranvec}[1]{\bm{\uppercase{#1}}}

\newtheorem{definition}{Definition}[section]
\newtheorem{proposition}{Proposition}[section]
\newtheorem{theorem}{Theorem}[section]
\newtheorem{coro}{Corollary}[section]
\newtheorem{lemma}{Lemma}[section]
\newtheorem{remark}{Remark}[section]
\title{Bayesian Credibility for GLMs}

\author{Oscar Alberto Quijano Xacur\thanks{Corresponding author: 
oscar.quijano@use.startmail.com. The authors gratefully
   acknowledge the partial financial support of NSERC grant 36860-2017.} ~and Jos\'e Garrido
  \\
  Concordia University, Montreal, Canada}

\allowdisplaybreaks
\begin{document}
\maketitle

\begin{dedication}
\hspace{3cm}{Dedicated to the memory of Bent J{\o}rgensen;}\\
\hspace{3cm}{a true scholar, a gentlemen and a great person.}
\end{dedication}

\abstract{We revisit the classical credibility results of Jewell
  \cite{jewell} and B\"uhlmann \cite{Buhlmann} to obtain credibility
  premiums for a GLM using a modern Bayesian approach. Here the prior
  distributions can be chosen without restrictions to be conjugate to
  the response distribution. It can even come from out--of--sample
  information if the actuary prefers.
 
  Then we use the relative entropy between the ``true'' and the
  estimated models as a loss function, without restricting credibility
  premiums to be linear.  A numerical illustration on real data shows
  the feasibility of the approach, now that computing power is cheap,
  and simulations software readily available.}



\section{Introduction}

The well known classical result from Jewell \cite{jewell} 
gives exact linear credibility estimators for the exponential family. More
precisely, it applies to random variables $Y\/$ with distribution in
some exponential dispersion family (EDF), i.e.~with density

\begin{equation}
  \label{eq:edf-density}
  f(y|\theta,\phi) = a(y,\phi)\exp\left(\frac{1}{\phi} \left\{y\theta
      - \kappa(\theta) \right\}\right),\qquad \theta\in\Theta, \phi\in\Phi\/.
\end{equation}
Assuming that $\phi\/$ is known, Jewell uses the following prior density on $\theta$: 
\begin{equation}
  \label{eq:jewell-prior}
  \pi_{n_0,x_0}(\theta) \propto \exp\big(n_0 [x_0 \theta - \kappa (\theta)]\big)\/, 
\end{equation}
for some hyper--parameters $n_0>0\/$ and $x_0\/$ in the
support of $Y\/$. For a conditionally i.i.d. sample $y_1,\ldots,y_n\/$ of $Y\/$,
given $\theta\/$, Jewell showed that the marginal mean of $Y\/$ given this sample is
\begin{align}
  \label{eq:linear-credibility-expression}
  \esp[Y | y_1\ldots y_n] &= \frac{\phi n_0}{\phi n_0 + n} x_0 + \frac{n}{\phi n_0 + n} \bar{y} \\
  &= (1-z) x_0 + z \bar{y}\/, \nonumber
\end{align}
where $\bar{y}=\frac{1}{n}\sum_{i=1}^n y_i\/$ and $z=\frac{n}{\phi n_0 +
  n}\/$ is known as the credibility factor.  Note that
\eqref{eq:linear-credibility-expression} is the B\"uhlmann \cite{Buhlmann} Bayesian point
estimator of the mean of $Y\/$ that minimizes the expected square loss
error and it is known as the linear credibility estimator.

The purpose of this article is to present a methodology for obtaining
credibility estimators that generalize these ideas and allow them to
be applicable to Generalized Linear Models (GLMs).

Hachemeister \cite{hachemeister} and De Vylder \cite{devylder} give
classical credibility results for regression models. The main idea
in these articles is similar to B\"uhlman's: they first impose linearity
on the regression covariates and then find the optimal linear
parameter estimators by minimizing a distance or error function.

Nelder and Verall \cite{nelder-verall-credibility} and Ohlsson \cite{ohlsson-credibility}
propose linear credibility estimators for GLMs. Although different in
substance, these two articles have in common that they are both
likelihood based and both rely on random effects in order to
obtain credibility estimates. The resulting Generalized Linear Mixed Models (GLMMs) credibility premiums remain essentially linear, which differs fundamentally from the method proposed here. 

In this article we use a modern Bayesian approach to obtain credibility
estimators. We focus on extending Jewell's results to Generalized Linear
Models (GLMs). Our estimator is not linear as we do not impose linearity constraints, nor force the prior to be conjugate. It also offers the advantage of considering the uncertainty on the dispersion parameter.

The exposition is divided as follows. In Section 2 we review the
essential elements needed to introduce our estimator. Section 3 
presents entropic estimators and some properties of the
unit deviance that are important for the rest of the article. In
Section 4 we discuss the un--feasibility of exact linear credibility for
GLMs. Section 5 proves that it is possible to obtain entropic
estimators for GLMs, which is our proposed credibility
approach. Moreover, Proposition \ref{prop:entropic-beta} suggests an
algorithm to obtain the estimators. 

Finally, in Section 6 we show that our
method is applicable to real life datasets by fitting the model to a
publicly available dataset. A commented version of the R code used for
this can be found at
\url{https://gitlab.com/oquijano/bayesian-credibility-glms}.



\section{Preliminaries}

\subsection{Exponential Dispersion Families}
In \eqref{eq:edf-density} $\theta\/$ and $\Theta\/$ are called the
canonical parameter and canonical space, respectively and $\phi\/$ is
known as the dispersion parameter. A neat property of reproductive
exponential families is that for $\theta\in\mbox{int} \left(
  \Theta\right)\/$ (here int stands for interior),
\begin{equation}
  \esp[Y] = \dot{\kappa}\left(\theta\right)\qquad \mbox{and}\qquad  \var[Y]=\phi\ddot{\kappa}\left(\theta\right)\/,
  \label{eq:kappa-derivatives}
\end{equation}
where $\dot{\kappa}=\kappa'\/$ and
$\ddot{\kappa}=\dot{\kappa}'\/$. This motivates the following
definitions.

\begin{definition}
  Given an exponential dispersion family, the mean domain of the
  family is defined as
  \begin{equation*}
    \Omega = \left\{\mu = \dot{\kappa}\left(\theta\right) : \theta \in \textrm{int}\left(\Theta\right) \right\}\/.
  \end{equation*}
\end{definition}

Another important property is that the support of the distribution
only depends on $\phi\/$ (and not on $\theta\/$). For a given family, let
$C_\phi\/$ be the convex support of any member of the family with
dispersion parameter $\phi\/$. We define the convex support of the
family as
\begin{equation*}
  C_\Phi=\bigcup_{\phi\in\Phi} C_\phi\/.
\end{equation*}

\begin{definition}
  The unit deviance function of an exponential dispersion family  is
  defined as $d: C_\Phi\times \Omega \rightarrow [0,\infty)\/$ with
  \begin{equation}
    \label{eq:unit-deviance}
    d\left(y,\mu\right) = 2 \left[ \sup_{\theta\in\Theta}\{\theta y - \kappa (\theta)\} -
      y \dot{\kappa}^{-1}(\mu) + \kappa\big(\dot{\kappa}^{-1}(\mu)\big)
 \right]\/.
  \end{equation}
\end{definition}

The unit deviance function plays an important role in the theory of
GLMs. The model assessment of a GLM is through hypothesis tests that
are based on the asymptotic behaviour of the unit deviance function.
It also allows to re-parametrize \eqref{eq:edf-density} as
\begin{equation}
  \label{eq:mean-value-par}
  f(y|\mu,\phi) = c(y,\phi)\exp\left(-\frac{1}{2\phi}d(y,\mu) \right)\/.
\end{equation}
This is known as the mean--value parameterization and it is the one 
used in this article.

When the canonical space $\Theta\/$ is open, it is said that the EDF
is regular. In this case $C_\Phi=\Omega$ and \eqref{eq:unit-deviance}
is equivalent to
\begin{equation}
  \label{eq:regular-unit-deviance}
      d\left(y,\mu\right) = 2 \left[ y\{ \dot{\kappa}^{-1}(y)
        -\dot{\kappa}^{-1}(\mu) \}  - 
        \kappa\big(\dot{\kappa}^{-1}(y)\big) +
        \kappa\big(\dot{\kappa}^{-1}(\mu)\big)
 \right]\/.
\end{equation}
In the rest of the article we work with regular
EDFs. Notice that all the usual distributions used for GLMs are
regular (e.g.~all the distributions in the Tweedie family are
regular).

There is a useful property of reproductive exponential dispersion
families that allows for data aggregation. J{\o}rgensen's notation (from
\cite{Jorgensen-book}) is very convenient to express this
property: given a fixed exponential family, if \(Y\) has mean \(\mu\)
and density given by \eqref{eq:mean-value-par}, we say that it is
\(ED(\mu,\phi/w)\/\) distributed. The property is then as follows: if
\(Y_1,Y_2,\cdots,Y_n\/\) are independent, and \(Y_i \sim
ED(\mu,\phi/w_i)\/\), then
\begin{equation}
  \label{eq:aggregate-equation}
  \bar{Y}=\frac{w_1Y_1+\cdots+w_nY_n}{w_+}\sim ED(\mu,\phi/w_+),\qquad
  w_+=\sum_{i=1}^n w_i\/.
\end{equation}

\subsubsection{A Note on Aggregating Discrete Exponential Dispersion Models }

There are two usual parametrizations of exponential dispersion
families. \eqref{eq:edf-density} gives the density of
\emph{reproductive} EDFs and it is used for continuous
distributions. Discrete distributions are usually parametrized as
\emph{additive} EDFs, whose densities have the form
\begin{equation}
  \label{eq:additive-edf}
    f(y|\theta,\phi) = a(y,\phi)\exp\left( y\theta
      -\frac{1}{\phi} \kappa(\theta) \right),\qquad \theta\in\Theta, \phi\in\Phi.
\end{equation}
Both parametrizations are defined and discussed in J{\o}rgensen
\cite{Jorgensen-book} and J{\o}rgensen \cite{jorgensen-deviance}.
GLMs assume the reproductive parameterization (see Nelder and Wedderburn
\cite{Neld:Wedd:1972}). Now, for many discrete EDFs, the dispersion
parameter has a known value. Specifically, for the Poisson, Bernoulli
and negative binomial distributions $\phi = 1\/$. This makes
\eqref{eq:edf-density} and \eqref{eq:additive-edf} the same
parametrization and allows such distributions to enter the GLMs
framework. Nevertheless, it is important to be aware that for the
discrete case, one cannot aggregate data using
\eqref{eq:aggregate-equation}. The properties of the Poisson
distribution allow to use an offset for data aggregation (see for
example Section 9.5 of Kaas, Goovaerts et al. \cite{modern-act}). Other discrete
distributions can be aggregated using quasi-likelihood.

\subsection{Relative Entropy}

Let $m_i\/$ be probability measures with $dm_i(\bm{y})=f_i(\bm{y})ds(\bm{y})\/$ for
some density functions $f_1\/$, $f_2\/$ and some probability measure $s\/$,
with $m_1\equiv m_2 \equiv s\/$ (that is $m_1\/$, $m_2\/$ and $s\/$ are absolutely continuous with respect to each other).

\begin{definition}
  The relative entropy of $m_2$ from $m_1$ is defined as
  \begin{equation*}
    D(m_1\parallel m_2) =
    \esp_{m_1}\left[\log\left(\frac{f_1(\bm{Y})}{f_2(\bm{Y})}\right) \right] =
    \int \log\left(\frac{f_1(\bm{y})}{f_2(\bm{y})}\right)dm_1(\bm{y})\/.
  \end{equation*}
\end{definition}

This definition was introduced by Kullback and Leibler in
\cite{kullback-leibler-article}. $D(\cdot \parallel \cdot)\/$ is often
called the Kullback--Leibler divergence, nevertheless we prefer the
term relative entropy since what they called divergence between
$m_1\/$ and $m_2\/$ was $D(m_1 \parallel m_2) + D(m_2 \parallel m_1)\/$.

The relative entropy is a measure of information. As such, it satisfies the
invariance property. Intuitively this means that bijective
transformations do not increase or decrease the information. 
The following paragraph expresses this formally.

Let $(\Omega_1,\sigmaf,m_i)$ and $(\Omega_2,\sigmag,\nu_i)\/$, for
$i=1,2\/$, be probability spaces and $T:\Omega_1\rightarrow \Omega_2\/$ a
measurable transformation such that $\nu_i(G)=m_i(T^{-1}(G))\/$, for
$G\in\sigmag\/$. Define also $\gamma(G)=s(T^{-1}(G))\/$. Since
$m_1\equiv m_2 \equiv s\/$, then $\nu_1\equiv \nu_2
\equiv \gamma\/$. This implies, by Radon--Nykodim's theorem that there
exist $g_1\/$ and $g_2\/$ such that
\begin{equation*}
  \nu_i(G)=\int_G g_i(\bm{y}) d\gamma(\bm{y}),\qquad G\in\sigmag\/.
\end{equation*}
With these definitions in mind, the following theorem asserts the
invariance property of the relative entropy. Its proof can be found in
Chapter 2 of Kullback \cite{kullback-book}.
\begin{theorem}
  $D(m_1 \parallel m_2)=D(\nu_1 \parallel \nu_2)\/$ if and only if $T\/$ is a
  bijective transformation. 
\end{theorem}

\subsection{GLMs}
In a GLM the response variable is assumed to follow a EDF with density
\begin{equation}
  \label{eq:exponential-density}
  f(y|\theta,\phi)=a(y,\phi)\exp\left(
    \frac{w}{\phi} 
    \{y\theta-\kappa( \theta )\}
  \right)\/,
\end{equation}
Note that $\phi\/$ in \eqref{eq:edf-density} corresponds to $\phi/w\/$
in \eqref{eq:exponential-density} which implies that the mean and
variance can be expressed as \(\mu=\kappa'(\theta)\/\) and
\(\sigma^2=\phi \kappa''(\theta)/w\/\), respectively. Here $w\geq 0\/$ is know
as the weight. In applications $w\/$ is known usually and $\phi\/$ needs
to be estimated. It is further assumed that there is a vector of
explanatory variables, also known as covariates,
\(\bm{x}=(x_1 \cdots x_p)^T\/\), a vector of coefficients
\(\bm{\beta}=(\beta_0~\beta_1 \cdots \beta_p)^T\/\) and a function \(g\/\)
known as the link function such that
\begin{equation}
  \label{eq:link-equation}
  g(\mu)=\beta_0+x_1\beta_1+\cdots+x_p\beta_p\/.
\end{equation}
It is useful for further developments to express the canonical
parameter \(\theta\/\) in terms of the coefficients. Since $\mu =
\kappa'(\theta) \equiv \dot{\kappa}(\theta)\/$ then:
\begin{align}
  \label{eq:theta-parameters}
  ( g \circ \dkappa ) (\theta) &= \beta_0+x_1\beta_1+\cdots+x_p\beta_p
                              \nonumber \\
  \theta&=( g \circ \dkappa )^{-1}(\beta_0+x_1\beta_1+\cdots+x_p\beta_p)\/.
\end{align}

The population can be divided into different classes according to the
values of the explanatory variables. Thus, given a sample, we can
group together all the observations that share the same values of
the explanatory variables and aggregate them using
\eqref{eq:aggregate-equation}. It is important to mention that with
this grouping there is no loss of information for estimating the mean
since $\bar{Y}\/$ is a sufficient statistic for $\theta\/$ (but not for
$\phi\/$, thus some information is lost for the estimation of $\phi\/$).

Possibly after aggregating, let \(m\/\) be the number of classes and
\(\bm{\theta}\in \Theta^m \), where \( \Theta^m =
\left\{(\theta_1 \cdots \theta_m)^T: \theta_1,\ldots,\theta_m\in\Theta
\right\}\) . The density of the sample can be expressed as
\begin{equation}
  \label{eq:glm-density}
  f(\bm{y} |\bm{\theta},\phi) =
  A(\bm{y},\phi) \exp \left( 
    \frac
    {\bm{y}^TW\bm{\theta} - \bm{1}^T W \bm{\kappa} (\bm{\theta})}
    {\phi}
  \right)\/,\qquad \bm{y}\in\realn^m\/,
\end{equation}
where 
\(\bm{\kappa}(\bm{\theta})=\big(\kappa(\theta_1) \cdots \kappa(\theta_m)\big)^T\/\),
\(W=\mbox{diag}(w_1,\cdots,w_m)\/\), with \(w_i\/\) being the sum of all the 
weights in the \(i\)-th class,
\(\bm{1}=(1 \cdots 1)^T\/\) and \(A(\bm{y},\phi) = \prod_{i=1}^m \big(a(y_i,
\frac{w_i}{\phi})\big)\/\). In order to express \(\bm{\theta}\/\) in terms of 
\(\bm{\beta}\/\), we define the following maps
\begin{equation*}
\bm{\mu}=\bm{\dkappa}(\bm{\theta})=\left(
  \begin{array}{c}
    \dkappa(\theta_1)\\ \vdots \\ \dkappa(\theta_m)
  \end{array}
\right)\/,\quad
G(\bm{\mu})=G\left( 
  \begin{array}{c}
    \mu_1\\ \vdots \\ \mu_m
  \end{array}
\right)=\left(
  \begin{array}{c}
    g(\mu_1)\\ \vdots \\ g(\mu_m)
  \end{array}  
  \right)\/,  
\end{equation*}
and the design matrix
\begin{equation*}
  X=\left(
    \begin{array}{cc}
      1 & \bm{x_1}^T\\
      \multicolumn{2}{c}{\vdots}\\
      1 & \bm{x_m}^T
    \end{array}
\right)\/,
\end{equation*}
where \(\bm{x}_i\/\) is the vector of explanatory variables for the
\(i\)-th class. Along this article we assume that $X$ has full rank
and that the model is not saturated (i.e. $p+1<m$). With all these
definitions, we have that
\begin{align}
  \label{eq:vector-theta-parameter}
  G(\bm{\mu})&=X\bm{\beta}, \nonumber\\
  (G\circ \bm{\dkappa})(\bm{\theta})&=X\bm{\beta}, \nonumber\\
  \bm{\theta} &= (G\circ \bm{\dkappa})^{-1}(X\bm{\beta})\/.
\end{align}

It is useful to reparameterize \eqref{eq:glm-density} in terms of the
mean vector $\bm{\mu}\/$ instead of $\bm{\theta}\/$. Using the mean value
parameterization (this is \eqref{eq:mean-value-par} but substituting
$\phi\/$ for $\phi/w\/$), \eqref{eq:glm-density} can be reparameterized as
\begin{equation}
  \label{eq:glm-deviance-par}
  f(\bm{y}|\bm{\mu},\phi) = C(\bm{y},\phi)\exp \left(-\frac{1}{2\phi}D(\bm{y},\bm{\mu}) \right)\/,
\end{equation}
where $C(\bm{y},\phi)=\prod_{i=1}^mc(y_i,\frac{\phi}{w_i})\/$, and
$D:\Omega^m \times \Omega^m \rightarrow [0,\infty) $ with
\begin{equation}
  \label{eq:glm-deviance-def}
  D(\bm{y},\bm{\mu})=\sum_{i=1}^mw_id(y_i,\mu_i)\/,
\end{equation}
where $\Omega^m=\left\{(\mu_1 \cdots \mu_m)^T: \mu_1,\ldots,\mu_m \in
  \Omega \right\}$. $D\/$ is called the deviance of the model. We give
here some of its properties:
\begin{itemize}
\item Given a sample, finding the mle of $\bm{\theta}\/$ is equivalent
  to finding the value of $\bm{\beta}\/$ that minimizes the deviance.
\item $D\/$ can be used to estimate the dispersion parameter (although
  it is not the only method). The deviance estimator of $\phi\/$ is
  given by
  \begin{equation*}
    \hat{\phi}=\frac{D(\bm{y},\hat{\bm{\mu}})}{m-p}\/.
  \end{equation*}
\item The asymptotic distribution of $D\/$ plays an important role in
  model assessment and variable selection.
\end{itemize}
For further details about the use and properties of the deviance we
recommend J{\o}rgensen \cite{jorgensen-deviance}.

\section{Entropic Estimator}


A posterior distribution is more informative than a point estimation
since it reflects our uncertainty about the true parameter. Now, in
insurance, it is necessary to charge a premium, which is
a point estimate. In this section we define the
point estimators that we propose as credibility premiums.

Consider a parametric family of distributions with a parameter
$\bm{\theta}$ to be estimated. Assume that $\bm{\theta}_0$ is the ``true''
parameter. In Bayesian point estimation one first chooses a loss
function $\lossf(\bm{\theta}_0,\bm{\theta}_1)$ that represents the cost of
estimating $\bm{\theta}$ to be $\bm{\theta}_1$ instead of $\bm{\theta}_0$. Now, since
$\bm{\theta}_0$ is not known, we define a risk function as
\begin{equation*}
  \riskf(\theta)=\esp\left[ \lossf(\bm{\theta}_0,\bm{\theta}) \right],
\end{equation*}
where the expectation is taken with respect to the posterior
distribution of $\bm{\theta}$. Then the point estimator $\hat{\bm{\theta}}$
of $\theta_0\/$ is the value of $\theta\/$ that minimizes
$\riskf$. That is:
\begin{equation*}
  \hat{\bm{\theta}} = \argmin_\theta \riskf(\bm{\theta})\/.
\end{equation*}

The entropic estimator is defined as the Bayesian point estimator when
the loss function $\lossf$ is the relative entropy of the
distribution with the real parameter $\bm{\theta_0}$ over the estimated
one.

More precisely, assume that the density of a random vector $\bm{Y}$
depends on a parameter $\bm{\theta}$. Denote with $f(\bm{Y}|\bm{\theta}_1)$ the
density of $\bm{Y}$ when $\bm{\theta}$ takes some value $\bm{\theta}_1$. The loss
function is defined as


\begin{equation*}
  \lossf(\bm{\theta}_0,\bm{\theta}) = \esp_{\bm{\theta}_0}\left[\log\left(\frac{f(\bm{Y}|\bm{\theta}_0)}{f(\bm{Y}|\bm{\theta})}\right) \right].
\end{equation*}
The corresponding risk function is then defined as
\begin{equation*}
  \riskf(\bm{\theta}) = \esp\left[ \log\left( \frac{f(\bm{Y}|\bm{\theta}_0)}{f(\bm{Y}|\bm{\theta})}  \right) \right] 
   := \esp_\pi \left[ \esp_{\bm{\theta}_0}\left[ \log\left( \frac{f(\bm{Y}|\bm{\theta}_0)}{f(\bm{Y}|\bm{\theta})}  \right) \right] \right],
\end{equation*}
where $\esp_\pi\/$ is the expectation taken with respect to the
posterior distribution of $\bm{\theta}\/$.

\begin{definition}
  The entropic estimator is defined as
  \begin{equation*}
    \hat{\bm{\theta}} = \argmin_{\bm{\theta}} \esp\left[\log\left(\frac{f(\bm{Y}|\bm{\theta}_0)}{f(\bm{Y}|\bm{\theta})}\right) \right]\/,
  \end{equation*}
  where the expectation is taken with respect to the posterior
  distribution of $\bm{\theta}\/$.
\end{definition}

Consider $\bm{\theta}$ to be the parameter of some model and let
$\bm{\beta}=g(\bm{\theta})$ be a bijective transformation. Entropic
estimators have the appealing invariance property that if $\bm{\theta}^*$ is the
entropic estimator of $\bm{\theta}$, then $\bm{\beta}^*=g(\bm{\theta}^*)$ is the
entropic estimator of $\bm{\beta}$. This property is called 
\emph{invariance} (this terminology is consistent with Bernardo \cite{bernardo}) .

Not all estimators have this property. For example the estimator that
minimizes the square loss error, i.e.~the posterior mean, is not
invariant. On the other hand, it is well known that maximum likelihood
estimators are invariant.

\subsection{Entropic Estimators for univariate EDFs}
A key part of this article is to show how to find entropic estimators
for GLMs. This is done in Section 4. In the rest of this section, we
focus on entropic estimators for univariate EDFs and their relation to
linear credibility.  The main result of the section is Proposition
\ref{prop:eq-pm-ee}, which is preceded by two technical lemmas that
show properties of the unit deviance that are fundamental for finding
entropic estimators of exponential families and GLMs (see the appendix
for the proofs of these lemmas).

\begin{lemma}
  \label{lem:unit-deviance-decomposition}
  Let $d\/$ be the unit deviance of a univariate EDF in
  \eqref{eq:regular-unit-deviance}. Then, there exist functions $d_1\/$ and
  $d_2\/$ such that for $(y,\mu)\in \Omega\times \Omega\/$, we have
the following decomposition:
  \begin{equation}
    \label{eq:unit-deviance-decomposition}
    d(y,\mu) = d_1(y) + d_2(y,\mu)\/.
  \end{equation}
  Moreover, $d_2\/$ has the property that if $Y\/$ is a random variable
  with support in $\Omega\/$, and $\mu\/$ is fixed, then
  \begin{equation*}
    \esp\big[d_2(Y,\mu)\big] = d_2\big(\esp[Y],\mu\big)\/.
  \end{equation*}
\end{lemma}

\begin{lemma}
  \label{lem:d_2-minimization}
  Let $y\/$ be fixed, then
  \begin{enumerate}
  \item the value of $\mu\/$ that minimizes $d_2(y,\mu)\/$ is the same one
    that minimizes $d(y,\mu)\/$.
  \item $d_2(y,\mu)\/$ is minimized when $\mu=y\/$.
  \end{enumerate}
\end{lemma}

Table \ref{tab:deviance-decomposition} shows $d$, $d_1$ and $d_2$ for
the normal, Poisson and gamma distributions.

\begin{table}[h]
  \centering
{\footnotesize
\begin{tabular}{|c|c|c|c|}
  \hline
  Distribution & $d(y,\mu)$ & $d_1(y)$ & $d_2(y,\mu)$ \\  
  \hline
  Normal & $(y-\mu)^2$ & $y^2$ & $\mu^2 - 2y\mu$\\
  \hline
  Poisson & $2\left\{y\log\left(\frac{y}{\mu}\right) - (y-\mu) \right\}$ & $2y[\log(y)-1]$&$2\left[\mu-y\log(\mu) \right]$ \\
  \hline
  Gamma & $2\left\{ \log\left(\frac{\mu}{y}\right) + \frac{y}{\mu}-1  \right\}$ & $2\left[\frac{y}{\mu} + \log(\mu)\right] $ & $2\left[\frac{y}{\mu}+\log\left(\frac{\mu}{y}\right) -1\right]$ \\
  \hline
\end{tabular}}
  \caption{Deviance decomposition of some common EDF's}
  \label{tab:deviance-decomposition}
\end{table}

\begin{proposition}
  \label{prop:eq-pm-ee}
  Let $Y$ be a random variable whose density is given by
  \eqref{eq:mean-value-par} for some unknown values of $\mu$ and
  $\phi$, $\pi(\mu,\phi)\/$ be a prior distribution for
  $(\mu,\phi)$, the vector $\bm{y}=(y_1 \ldots y_n)^T\/$ be a
  conditionally i.i.d. sample given $(\mu,\phi)\/$, and
  $\pi(\mu,\phi|\bm{y})\/$ the corresponding posterior. The entropic
  estimator $\hat{\mu}\/$ of $\mu\/$ is then given by
  \begin{equation*}
    \hat{\mu} = \esp[Y|\bm{y}] = \esp_\pi\big[\esp_{\mu,\phi}[Y]\big]\/,
  \end{equation*}
  where $\esp_\pi\/$ represents the expectation with respect to the
  posterior distribution and $\esp_{\mu,\phi}\/$ represent the
  expectations with respect to fixed values of $(\mu,\phi)\/$.
\end{proposition}
\begin{proof}
Let $(\mu_0,\phi_0)\/$  be the true parameters. By Lemma 
\ref{lem:unit-deviance-decomposition}, the entropic risk measure can
be expressed as
\begin{align*}
  \riskf(\mu,\phi) &=\esp\left[
                     \log\left(\frac{f(Y|\mu_0,\phi_0)}{f(Y|\mu,\phi)}
                     \right) \right] \\
                   &=\esp\left[
                     \log\left(\frac
                     {c(Y,\phi_0)\exp\left(-\frac{1}{2\phi_0}d(Y,\mu_0)\right)}
                     {c(Y,\phi)\exp\left(-\frac{1}{2\phi}d(Y,\mu)\right)}
                     \right) \right] \\
                   &=\esp\left[
                     \log\left(c(Y,\phi_0)\exp\left(-\frac{1}{2\phi_0}d(Y,\mu_0)\right)\right)\right]\\ 
                   &\quad  - \esp[\log(c(Y,\phi))]
                    + \frac{1}{2\phi}\esp[d(Y,\mu)]\\
                   &=\esp\left[\log\left(c(Y,\phi_0)\exp\left(-\frac{1}{2\phi_0}d(Y,\mu_0)\right)\right)\right]\\ 
                   &\quad  - \esp[\log(c(Y,\phi))]
                     + \frac{1}{2\phi}\esp[d_1(Y)] + \frac{1}{2\phi}d_2(\esp[Y],\mu)\/.
\end{align*}
Note that, regardless of the value of $\phi\/$, the value of $\mu\/$
that minimizes the expression above is the same one that minimizes the
simpler function
\begin{equation*}
  \riskf_1(\mu) = d_2\big(\esp[Y],\mu\big)\/.
\end{equation*}
Then, by Part 2 of Lemma \ref{lem:d_2-minimization}, the entropic
estimator of $\mu\/$ is given by $\hat{\mu} = \esp[Y]\/$.
\end{proof}

This result shows that for univariate EDFs the posterior mean not only
minimizes the expected square error risk, but also the posterior
entropic risk. In the following sections we will see that this
property does not generalize to GLMs due to the difference of
dimension between the response vector and the regression
coefficients. For now, realize that a direct consequence of
Proposition \ref{prop:eq-pm-ee} is that Jewell's estimator
in \cite{jewell} is an entropic estimator.

\begin{coro}
  The linear credibility estimator
  \eqref{eq:linear-credibility-expression} is the entropic estimator
  when $\phi\/$ is assumed known and \eqref{eq:jewell-prior} is used as
  prior for $\theta\/$.
\end{coro}

\section{Linear Credibility for GLMs}

This section discusses whether it is possible to extend Jewell's
result to GLMs. In other words we address the question: is there a
prior for the regression coefficients $\bm{\beta}\/$ for which the
posterior mean is a weighted mean between an out--of--sample estimate
and the sample mean of a GLM?

There are two ways in which the question above can be interpreted. One
could think of it as all $m\/$ dimensions  having 
the same credibility factor, that is, the credibility premium
$\hat{\bm{\mu}}_c$ is given by
\begin{equation}
  \label{eq:glm-linear-credibility-type1}
  \hat{\bm{\mu}}_c = z \bar{\bm{y}}  + (1-z) \bm{M}\/,
\end{equation}
where $\bar{\bm{y}}$ is the GLM observed sample mean (i.e.~a vector for
which \eqref{eq:aggregate-equation} applies to each coordinate),
$\bm{M}\/$ is a vector of out--of--sample ``manual" premiums as
coordinates and $z\in (0,1)\/$ is the credibility factor. We call this
interpretation \emph{Linear Credibility of Type 1}.

The other interpretation is to give a different credibility factor to
each coordinate. This is
\begin{equation}
  \label{eq:glm-linear-credibility-type2}
  \hat{\bm{\mu}}_c = Z \bar{\bm{y}} + (I-Z) \bm{M}\/,
\end{equation}
where $\hat{\bm{\mu}}_c\/$, $\bar{\bm{y}}$ and $\bm{M}\/$ are as in
\eqref{eq:glm-linear-credibility-type1}, but $Z=\diag(z_1,\ldots,z_m)\/$,
where $z_i\/$ is the credibility factor of the $i$-th class and $I\/$ is
the identity matrix. We call this interpretation \emph{Linear
 Credibility of Type 2}. Note that linear credibility of Type 1 is
a special case of linear credibility of Type 2.

\subsection{Linear Credibility of Type 1 is Impossible}

Jewell's prior in \eqref{eq:jewell-prior} is a conjugate prior to
\eqref{eq:edf-density} that gives linear credibility premiums based 
on the posterior mean. Diaconis and Ylvisaker \cite{diaconis1979} 
generalized Jewell's result to multivariate exponential
dispersion models. Since a GLM assumes that the response vector
follows such a distribution, it could be conjectured that this
automatically implies linear credibility for GLMs. In what follows 
we show that this is not the case.

After adapting the conjugate prior discussed in \cite{diaconis1979} to
correspond to \eqref{eq:glm-density} so as to consider weights we get
\begin{equation}
  \label{eq:conjugate_prior_glm}
  \pi_{n_0,\bm{x_0}}(\bm{\theta})\propto
  \exp\big(n_0\{\bm{x_0}^T W\bm{\theta} - \bm{1}^T W\bm{k}(\bm{\theta}) \}\big)\mathbb{I}_{\Theta^m}(\bm{\theta})\/,
\end{equation}
where $n_0>0\/$, and $\bm{x_0}\in \Omega^m\/$ are the parameters of
the prior distribution, $\Theta^m =
\left\{(\theta_1 \ldots \theta_m)^T:\theta_1,\ldots,\theta_m \in \Theta
\right\}$, $\mathbb{I}_{\Theta^m}\/$ is an indicator function and
$\Omega^m=\left\{(\mu_1 \ldots \mu_m)^T: \mu_1,\ldots,\mu_m \in
  \Omega\right\}$. Theorem 3 of \cite{diaconis1979} proves that if the
support of \eqref{eq:glm-density} contains an interval then
\eqref{eq:conjugate_prior_glm} is the only prior that gives linear
credibility. This implies that for any continuous response (and also
for any Tweedie distribution), \eqref{eq:conjugate_prior_glm} is the
only prior that gives linear credibility. In the paper it is also
proven that \eqref{eq:conjugate_prior_glm} is the unique prior that
gives linear credibility for the binomial distribution and in Johnson
\cite{johnson1957} the same is proven for the Poisson distribution.

As shown in \eqref{eq:vector-theta-parameter}, $\bm{\theta}\/$ and
$\bm{\beta}\/$ are related by $\bm{\theta} = ((G\circ
\bm{\dkappa})^{-1}\circ X )(\bm{\beta})\/$. Thus, when a prior for
$\bm{\beta}$ is chosen, a distribution is induced on $\bm{\theta}$. In
what follows we refer to this distribution as the \emph{induced prior}
on $\bm{\theta}$. We have then that for continuous and Tweedie
distributions and for the Poisson and negative binomial, a prior on the
betas gives linear credibility if and only if the induced prior on
$\bm{\theta}$ is \eqref{eq:conjugate_prior_glm}.

Our strategy to prove the impossibility of linear credibility of type
1 is to show that no prior of $\bm{\beta}$ induces a prior on
$\bm{\theta}$ that has density \eqref{eq:conjugate_prior_glm}. We can
see that this is the case by focusing on the support of the induced
distribution. Since the support of \eqref{eq:conjugate_prior_glm} is
$\Theta^m$, then it is enough to prove that the support of every
induced prior of $\bm{\theta}$ is different than $\Theta^m$. 

\begin{proposition}
  \label{prop:proper-subset}
  For any prior of $\bm{\beta}$ in a non-saturated GLM, the support of
  the induced prior of $\bm{\theta}$ is a proper subset of $\Theta^m$.
\end{proposition}

\begin{proof}

  Since there is no restriction for the value of $\bm{\beta}$, it can
  take any value on $\realn^{p+1}$. This is represented on the left
  rectangle of Figure \ref{fig:diagrams-proper-subset-proof}.

  $X\bm{\beta}$ can take values in $R(X)$, where $R(X)$ is the range
  of $X$. Since $\dim(R(X))=p+1<m$, then $R(X)\subsetneq
  \realn^m$. This is represented in the middle rectangle of Figure
  \ref{fig:diagrams-proper-subset-proof}.

  Let $S$ be the support of the induced prior on $\bm{\theta}$. Then
  $S\subset (G\circ\bm{\dkappa})^{-1}(R(X)) :=\left\{
    (G\circ\bm{\dkappa})^{-1}(X\bm{\beta}):\bm{\beta}\in\realn^{p+1}
  \right\}$ (subset but not equality since some values of
  $\realn^{p+1}$ may not be in the support of $\bm{\beta}$). Now,
  $(G\circ\bm{\dkappa})^{-1}$ is a bijective function. Let $\bm{p}$ be
  a point in $\realn^m$ that is not in $R(X)$ and let
  $\bm{q}=(G\circ\bm{\dkappa})^{-1}(\bm{p})$. Then $\bm{q}\in\Theta^m$
  but $\bm{q}\in S$ because otherwise $(G\circ\bm{\dkappa})^{-1}$
  would not be one-to-one. This proves that
  $(G\circ\bm{\dkappa})^{-1}(R(X)) \subsetneq \Theta^m$ and therefore
  also that $S \subsetneq \Theta^m$. This is represented in the right
  rectangle of Figure \ref{fig:diagrams-proper-subset-proof}.

\begin{figure}[h]
  \centering
  \includegraphics[width=0.8\textwidth]{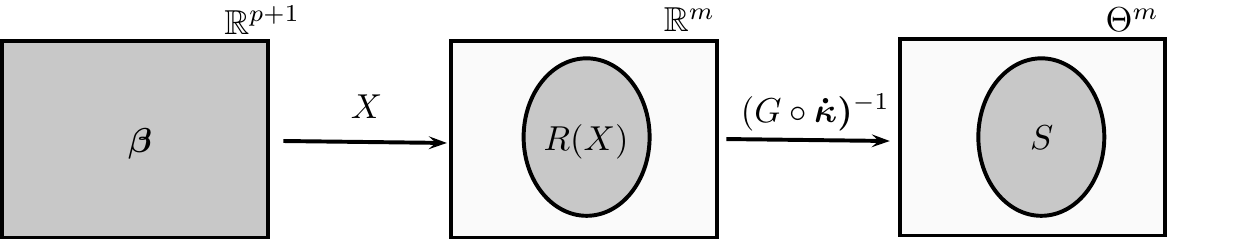}
  \caption{From left to right the grey zone represents the values
    that $\bm{\beta}$, $R(X)$ and $S$ can take, respectively.}
  \label{fig:diagrams-proper-subset-proof}
\end{figure}

\end{proof}
Now, on a different but related note, it is possible to generalize
\eqref{eq:conjugate_prior_glm} in a way that allows to obtain
conjugate priors that are suitable for GLMs. Define
\begin{equation*}
  \pi_1(\bm{\theta})\propto h(\bm{\theta})
  \exp\big(n_0\{\bm{x_0}^TW\bm{\theta} - \bm{1}^TW\bm{k}(\bm{\theta}) \}\big)\mathbb{I}_{\Theta^m}(\bm{\theta})\/,
\end{equation*}
where $h\/$ is some integrable function for which the integral on the
right hand--side above is finite and denote this distribution by
$D_{conj}(n_0,\bm{x_0})$. 

\begin{proposition}
  $\pi_1$ is a conjugate prior to \eqref{eq:glm-density} with
  posterior distribution
  $D_{conj}\left(n_0+\frac{1}{\phi},\frac{1}{n_0\phi +1}\bar{\bm{y}} +
    \frac{\phi n_0}{\phi n_0 + 1}\bm{x_0}\right)$.
\end{proposition}
\begin{proof}
  Let $\pi_1(\cdot|\bm{y})$ denote the posterior of $\pi_1$. Then, by
  definition of the posterior:
  \begin{align*}
    \pi_1(\cdot|\bm{y}) & \propto \pi_1(\bm{\theta})   f(\bm{y} |\bm{\theta},\phi) \\
    &\propto h(\bm{\theta})\exp\left(
      n_0\bm{x_0}^TW\bm{\theta} + \frac{\bm{y}^TW\bm{\theta}} {\phi}
      -n_0\bm{1}^TW\bm{k}(\bm{\theta}) - \frac{\bm{1}^T W \bm{\kappa}
        (\bm{\theta})}{\phi}\right)\\
    &= h(\bm{\theta}) \exp\left( (n_0\bm{x_0}^T +
      \frac{\bm{y}^T}{\phi}) W \bm{\theta} - \left( n_0 + \frac{1}{\phi}\right) W
      \bm{\kappa}(\bm{\theta}) \right) \\
    &= h(\bm{\theta}) \exp\left( 
      \left( n_0 + \frac{1}{\phi} \right) \left\{ 
        \frac{n_0\bm{x_0}^T+\frac{\bm{y}^T}{\phi}}{n_0+\frac{1}{\phi}}W\bm{\theta} -
        \bm{1}^T W \bm{\kappa}(\bm{\theta})
      \right\}
    \right)\\
    &= h(\bm{\theta}) \exp\left( 
      \left( n_0 + \frac{1}{\phi} \right) \left\{ 
        \frac{\phi n_0\bm{x_0}^T+\bm{y}^T}{\phi n_0+1}W\bm{\theta} -
        \bm{1}^T W \bm{\kappa}(\bm{\theta})
      \right\}
    \right),
  \end{align*}
  which proves the result.
\end{proof}

Now, in order for $\pi_1$ to overcome the problems that do not allow
$\pi\/$ in \eqref{eq:conjugate_prior_glm} to be used as a prior for
GLMs, it is only necessary to chose $h\/$ such that $\pi_1\/$ is
outside of $(G\circ\bm{\dkappa})^{-1}(R(X))\/$ with probability
zero. This way $\pi_1\/$ has the ``right'' support and there is a
distribution of $\bm{\beta}\/$ that gives this distribution when
transformed with $((G\circ \bm{\dkappa})^{-1}\circ X )\/$.

Two important remarks about $\pi_1\/$:
\begin{enumerate}
\item It does not give linear credibility (since this is impossible as
  has been shown above).
\item It is not easy to find an analytic expression for ${\bm{\mu}}\/$
  (although this might be possible for some choices of $\pi\/$). Thus
  most likely one has to use some numerical method or MCMC in order to
  find the posterior means, but this defeats the purpose of using a
  conjugate prior.
\end{enumerate}

\subsection{Linear Credibility of Type 2 is Sometimes Feasible}

Since the model is a GLM, there should be a $\hat{\bm{\beta}}_c\/$ such
that $\hat{\bm{\mu}}_c=G^{-1}(X \hat{\bm{\beta}}_c)\/$. Thus,
\eqref{eq:glm-linear-credibility-type2} becomes
\begin{equation}
  \label{eq:glm-linear-credibility-type2-beta}
  G^{-1}(X \hat{\bm{\beta}}_c) = Z \bar{\bm{y}} + (I-Z) \bm{M}\/.
\end{equation}
It turns out that for non saturated models (i.e.~$\dim(\bm{\beta})
<\dim(\bm{\mu})\/$), the existence of some $\hat{\bm{\beta}}_c\/$ for
which \eqref{eq:glm-linear-credibility-type2-beta} can be satisfied
depends on the observed sample. We demonstrate why this is the case
with a simple example in dimension 2.

Consider a situation in which you divide your population in only 2
segments using a binary covariate with no intercept (otherwise we
would have a saturated model). The design matrix in this case would be
\begin{equation*}
  X=\left(
    \begin{array}{c}
      0 \\
      1
    \end{array}
\right)\quad \mbox{ and } \quad  \hat{\beta}_c\in \realn\/.
\end{equation*}
Then, assuming a log--link function, the left hand side of
\eqref{eq:glm-linear-credibility-type2-beta} can be expressed as
\begin{equation*}
  \hat{\bm{\mu}}_c
    = G^{-1}(X \hat{\beta}_c) 
    = G^{-1} \left(     \begin{array}{c}
      0 \\
      \hat{\beta}_c
    \end{array}
  \right)
    = \left(
      \begin{array}{c}
       \exp (0) \\
      \exp (\hat{\beta}_c)
    \end{array}
\right)
    = \left(
      \begin{array}{c}
       1 \\
      \exp (\hat{\beta}_c)
    \end{array}
\right)\/.
\end{equation*}

If we graphed it we would see that the left hand side of
\eqref{eq:glm-linear-credibility-type2-beta} takes values 
only on the half upper side of the vertical line $x=1\/$.

Imagine now two scenarios. In Scenario 1, $ \bar{\bm{y}}  = (0.5,2)\/$
and $\bm{M} = (2,3)\/$, while in Scenario 2, $ \bar{\bm{y}} = (2,3)\/$ 
and $\bm{M} = (4,5)\/$.

\begin{figure}[h]
  \centering
  \begin{subfigure}[b]{0.495\textwidth}
    \includegraphics[width=\textwidth]{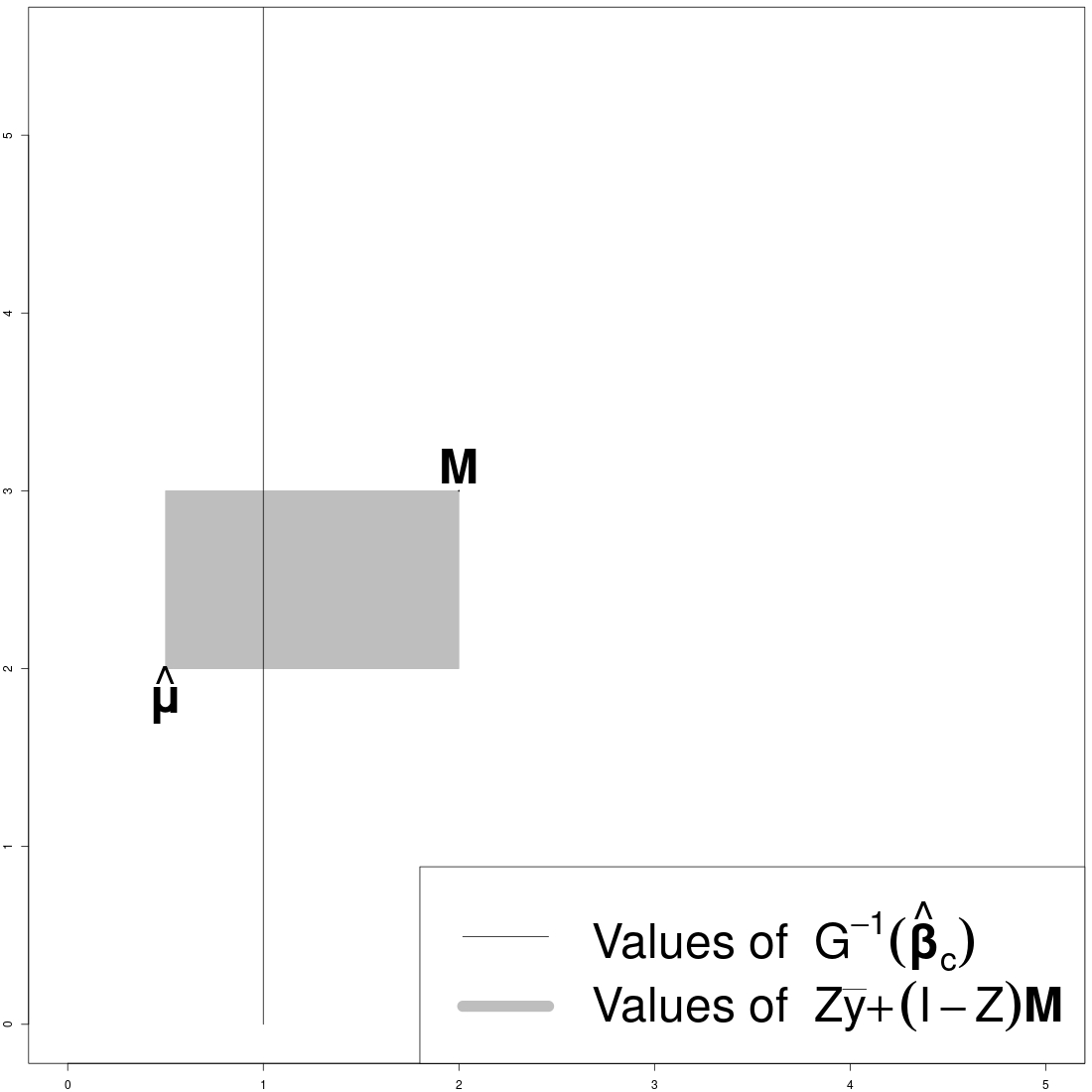}
    \caption{Scenario 1}
    \label{fig:scenario1}
  \end{subfigure}  
  \begin{subfigure}[b]{0.495\textwidth}
    \includegraphics[width=\textwidth]{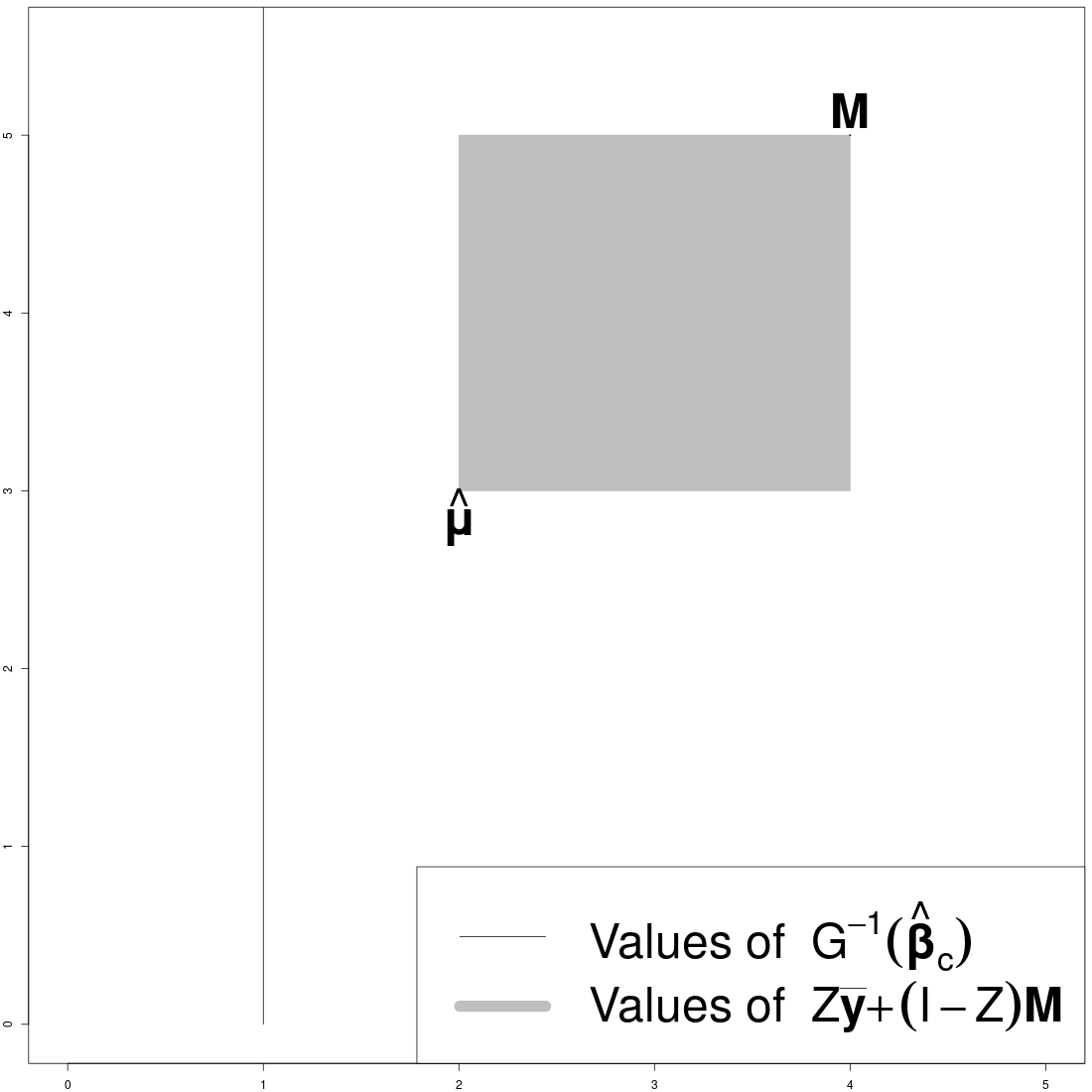}
    \caption{Scenario 2}
    \label{fig:scenario2}
  \end{subfigure}
  \caption{Values of the left and right hand side of
    \eqref{eq:glm-linear-credibility-type2-beta} in both scenarios}
  \label{fig:example-scenarios}
\end{figure}

As the values of the elements of $Z\/$ vary, the right hand side of
\eqref{eq:glm-linear-credibility-type2} can take the values of the
rectangle defined by $\bar{\bm{y}}$ and $\bm{M}\/$. Figure
\ref{fig:example-scenarios} shows graphs with the possible values of
the left and right hand side of
\eqref{eq:glm-linear-credibility-type2} for each scenario.

In both graphs, the vertical line represents the values of
$\hat{\bm{\mu}}_c\/$. The rectangle represents all the possible values
that $Z \bar{\bm{y}} + (I-Z) \bm{M}$ can take as the entries in the
diagonal of $Z$ vary from 0 to 1. In order to have exact linear
credibility of type 2, it is necessary for the line and the rectangle
to intersect. This is because the points of intersection, correspond
to combinations of values of $\bm{\beta}_c\/$ and $Z$ for which
\eqref{eq:glm-linear-credibility-type2-beta} holds. If there is no
intersection it is not possible to have linear credibility of type 2.

The graph of Scenario 1 shows that \eqref{eq:glm-linear-credibility-type2} 
is satisfied for some values of $Z\/$, while in the graph for
Scenario 2 it is impossible to satisfy
\eqref{eq:glm-linear-credibility-type2}.

The results of this section show that Jewell's result cannot be
generalized to GLM's. That is, no prior for the parameters of a GLM
guarantee linear credibility for all observed samples. In the next section we propose
credibility estimators for GLMs that are not linear.

\section{Entropic Credibility for GLMs}

The Bayesian models proposed by B\"uhlmann and Jewell, using conjugate
priors that ensured linear credibility formulas, made great sense in
the 1960's and 70's, when computational issues ruled out more general
(non--linear) Bayesian solutions. However, computing power is no
longer scarce nor expensive. In this section we propose a modern,
computational Bayesian approach to credibility.

\subsection{Estimation of the Mean \label{sec:mean}}

We propose an entropic estimator of the mean vector of a GLM as the
credibility premium. This section focuses on how to find such an
estimator. We start by enunciating the following technical lemma,
which is an extension of Lemma \ref{lem:unit-deviance-decomposition}
to greater dimensions. A proof can be found in the Appendix.

\begin{lemma} 
  \label{lem:glm-deviance-decomposition}
  Let $D\/$ be the deviance of a GLM (see
  \eqref{eq:glm-deviance-def}). Then there exist functions $D_1\/$ and $D_2\/$
  such that for $(\bm{y},\bm{\mu})\in \Omega^m \times \Omega^m\/$
  \begin{equation}
    \label{eq:deviance-decomposition}
    D(\bm{y},\bm{\mu})=D_1(\bm{y})+D_2(\bm{y},\bm{\mu})\/.
  \end{equation}
  Moreover, $D_2\/$ has the property that if $\ranvec{y}$ is a random vector
  with support in $\Omega^m\/$ and $\bm{\mu}\in\Omega^m\/$ is fixed, then
  \begin{equation}
    \label{eq:d2-property}
    \esp[D_2(\ranvec{y},\bm{\mu})] = D_2(\esp[\ranvec{y}],\bm{\mu})\/.
  \end{equation}
\end{lemma}
\noindent
In what follows, an arbitrary prior \(\pi\/\) is assumed (not necessarily 
conjugate) with posterior
\begin{equation}
  \label{eq:posterior-proportional-density}
  \pi( \bm{\beta},\phi | \bm{y}) \propto f(\bm{y} | \bm{\beta},\phi)\pi(\bm{\beta},\phi)\/,
\end{equation}
where \(f\/\) is as in \eqref{eq:glm-density} or, equivalently
\eqref{eq:glm-deviance-par}, depending on the chosen parameterization. 
As stated, \(\esp_\pi[\cdot]\/\) denotes expectation with respect to the 
posterior measure. Whenever the expectation symbol is used without a subindex, 
it means expectation with respect to the predictive posterior
distribution,
i.e.~\(\esp[\cdot] = \esp_\pi[\esp_{\bm{\beta},\phi}(\cdot)]\/\), where
\(\esp_{\bm{\beta},\phi}(\cdot)\/\) means expectation with respect to
the density in \eqref{eq:glm-density} with fixed coefficients vector
\(\bm{\beta}\/\) and fixed dispersion parameter \(\phi\/\).

\begin{proposition}
  \label{prop:entropic-beta}
  The entropic estimator $\bm{\beta}^*\/$ of the coefficients of a
  Bayesian GLM are equal to the maximum likelihood estimator of a
  frequentist GLM with the same covariates, response distribution and
  weights, but with an observed response vector equal to $\esp[\ranvec{y}]$.
\end{proposition}

\begin{proof}
  Let $(\bm{\beta}_0,\phi_0)\/$ be the real parameters and
  $(\bm{\beta},\phi)\/$ some fixed values. We use here for $f\/$
the mean value parameterization in \eqref{eq:glm-deviance-par}. Then,
  the risk function is given by
\begin{equation*}
  \riskf(\bm{\beta},\phi) = \esp_\pi\{\lossf[(\bm{\beta}_0,\phi_0),
  (\bm{\beta},\phi)]\}
  =\esp \left[ \log\left( \frac{f(\ranvec{y}|\bm{\mu}_0,\phi_0)}{f(\ranvec{y}|\bm{\mu},\phi)} \right) \right]\/,
\end{equation*}
where $\bm{\mu} = G^{-1}(X\bm{\beta})\/$ and $\bm{\mu}_0 =
G^{-1}(X\bm{\beta_0})\/$. Then, by Lemma
\ref{lem:glm-deviance-decomposition} and \eqref{eq:glm-deviance-par} the expression above becomes
\begin{align}
  \label{eq:glm-risk-function}
  \riskf (\bm{\beta},\phi) &= \esp\left[ \log(C(\ranvec{y},\phi_0) ) -
  \frac{1}{2\phi_0}D(\ranvec{y},\bm{\mu}_0)  \right] - \esp[ \log(C(\ranvec{y},\phi))] 
	\nonumber \\
&\quad+ \frac{1}{2\phi}\esp[D(\ranvec{y},\bm{\mu})] \nonumber \\
                           & = \esp \left[ \log(C(\ranvec{y},\phi_0) ) -
  \frac{1}{2\phi_0}D(\ranvec{y},\bm{\mu}_0)  \right] \nonumber \\
&\quad - \esp[ \log(C(\ranvec{y},\phi) )] +
\frac{1}{2\phi}\esp [D_1(\ranvec{y})+D_2(\ranvec{y},\bm{\bm{\mu}}) ] \nonumber \\
                           & = \esp\left[ \log(C(\ranvec{y},\phi_0) ) -
                             \frac{1}{2\phi_0}D(\ranvec{y},\bm{\mu}_0) +
                             \frac{1}{2\phi}D_1(\ranvec{y}) \right]\nonumber \\
&\quad -\esp[\log(C(\ranvec{y},\phi))] +\frac{1}{2\phi}
\esp[D_2(\ranvec{y},\bm{\mu})]\nonumber \\
                           & = \esp\left[ \log(C(\ranvec{y},\phi_0) ) -
                             \frac{1}{2\phi_0}D(\ranvec{y},\bm{\mu}_0) +
                             \frac{1}{2\phi}D_1(\ranvec{y}) \right] \nonumber \\
&\quad -\esp[\log(C(\ranvec{y},\phi))] + \frac{1}{2\phi} D_2(\esp[\ranvec{y}],\bm{\mu})\/.
\end{align}
The Bayesian point--estimator of $(\bm{\beta}_0,\phi_0)\/$ is given by
the vector $(\bm{\beta}^*,\phi^*)\/$ that minimizes $\riskf\/$. Let us
first focus on finding $\bm{\beta}^*\/$. Note that this is
equivalent to minimizing
\begin{equation}
  \label{eq:equiv-min}
  \riskf_1(\bm{\beta}) = D_2(\esp[\ranvec{y}],\bm{\mu})\/.
\end{equation}
We first need to compute $\esp[\ranvec{y}]\/$, which can be expressed as
\begin{equation*}
  \esp[\ranvec{Y}] = \esp_\pi \big[ \esp_{\bm{\beta}_0,\phi_0}[\ranvec{y}]  \big] =
  \esp_\pi\left[ G^{-1}(X\bm{\beta})\right]\/.
\end{equation*}
Now, \eqref{eq:posterior-proportional-density} gives $\pi(\cdot |
\bm{y})\/$ up to a normalizing constant, thus the expectation above
can be calculated with MCMC methods. In this way we consider the
problem of computing $\esp[\ranvec{y}]\/$ solved.

Compare now the minimization of $\riskf_1(\bm{\beta})\/$ with a different
optimization problem for which the solution method is well
known. Consider a frequentist (non--Bayesian) GLM with the same
response distribution, explanatory variables and weights. Imagine a
sample under this model in which the observed response vector is equal to
$\esp[\ranvec{y}]\/$. Using the mean value parameterization and Lemma
\ref{lem:glm-deviance-decomposition}, the log--likelihood function based on such
a sample is given by
\begin{align*}
  \ell(\bm{\beta},\phi) & = \log(C(\esp[\ranvec{y}],\phi)) -
  \frac{1}{2\phi} D(\esp[\ranvec{y}],\bm{\mu})\\
                        &= \log(C(\esp[\ranvec{y}],\phi)) -
                          \frac{1}{2\phi} D_1(\esp[\ranvec{y}]) -
                          \frac{1}{2\phi} D_2(\esp[\ranvec{y}],\bm{\mu})\/,
\end{align*}
where $\bm{\mu}=G^{-1}(X\bm{\beta})\/$. Since the only term that depends
on $\bm{\beta}\/$ is the third one, then maximizing
$\ell(\bm{\beta},\phi)\/$ is equivalent to minimizing
$D_2(\esp[\ranvec{y}],\bm{\mu})\/$, i.e.~the same as minimizing
$\riskf_1(\bm{\beta})\/$. Hence, by obtaining the mle of the regression
coefficients of this hypothetical frequentist GLM, we  obtain
$\bm{\beta}^*\/$ (or conclude that there is no solution, whenever this is the
case).
\end{proof}

Once $\bm{\beta}^*\/$ has been found, the invariance property of the
relative entropy allows to find the entropic premium
straightforwardly.

\begin{coro}
  \label{cor:entropic-mean}
  If $\bm{\beta}^*\/$ is the entropic estimator of the coefficients of a
  Bayesian GLM, then the entropic premium is given by
  \begin{equation}
    \label{eq:entropic-premium}
    \bm{\mu}^*=G^{-1}(X \bm{\beta}^*)\/.
  \end{equation}
\end{coro}

\begin{remark}
  \label{remark:saturated-model-entropic-premium}
  For a saturated model, i.e.~when the dimension of $\bm{\beta}\/$ is
  equal to the dimension of $\ranvec{y}$ (in other words $m=p+1\/$), the entropic
  premium is equal to $\esp[\ranvec{y}]$. This is because in a saturated model,
  the predicted mean is equal to the observed response mean.
\end{remark}

\subsection{Estimation of the Dispersion Parameter}

It is important to remark that the credibility estimator from the
previous section takes into consideration the uncertainty of the
dispersion parameter. This is the case because the posterior
distribution of $\bm{\beta}\/$ depends on the posterior of $\phi\/$.

This differs from classical credibility results where the
dispersion parameter is considered known (e.g.~Jewell \cite{jewell} and
Diaconis and Ylvisaker \cite{diaconis1979}). To the best of our knowledge there is only one
article that considers a prior distribution for the dispersion parameter,
Landsman and Makov \cite{landsman-makov}, about which we have the 
following remarks:

\begin{enumerate}
\item The exponential distribution for the index parameter is
  justified using the principle of maximum entropy. The authors 
  maximize the continuous entropy (that is entropy for continuous random
  variables), and use it for the index parameter assuming a known mean. Now, the
  continuous entropy does not have good properties as a measure of
  information. For instance it is not invariant under bijective
  transformations, which implies that one can loose or gain
  information by just transforming a random variable. Thus, the
  principle of maximum entropy is not a valid justification for the
  exponential distribution. Nevertheless, it is a valid prior and one
  can use it in those cases where it reflects properly the out of
  sample information.

\item A more serious problem exists with their result in Theorem 2;
  the integrals in (7) are carried out assuming that $\lambda\/$ is
  exponential with mean $\lambda_0\/$. In other words, these are
  computed assuming the prior distribution for $\lambda\/$. This is
  erroneous since it is the posterior distribution that should be used
  in this integral. This would be justified if the prior for
  $\lambda\/$ were natural conjugate. In this way the posterior of
  $\lambda\/$ would also be exponential, but the parameter of the
  posterior would be different than the parameter of the prior, in
  this case.

\end{enumerate}

We have not found a general procedure for obtaining the entropic
estimator of the dispersion parameter. We discuss here the cases for
which it can be found and present the difficulties in obtaining
a general solution. Notice that a point--estimator for $\phi\/$ is not
necessary to obtain the credibility premium (as seen in Section
\ref{sec:mean}) or its uncertainty (which is measured by the posterior
distribution).

Suppose that the credibility premium $\bm{\mu}^*\/$ has been
obtained. From \eqref{eq:glm-risk-function}, one can see that finding the
entropic estimator $\phi^*\/$ of the dispersion parameter is equivalent
to minimizing
\begin{equation}
  \label{eq:equivalence-minimize-phi}
  \riskf_2(\phi) = -\esp\left[ \log(C(\ranvec{y},\phi)) \right] + 
  \frac{1}{2\phi}\esp[D(\ranvec{y},\bm{\mu}^*)]\/,
\end{equation}
where $\bm{\mu}^*=G^{-1}(X\bm{\beta}^*)\/$. There are standard methods for
minimizing univariate functions, but $\riskf_2\/$ is more difficult because
the first expectation in \eqref{eq:equivalence-minimize-phi} depends
on $\phi\/$. We consider first a special case where this minimization is
rather straightforward. This is when there exists a function
$H:\realn^m\times \realn \rightarrow \realn\/$ such that
\begin{equation}
  \label{eq:simplified-phi-star}
  -\esp[\log(C(\ranvec{y},\phi))] = H(\esp[\ranvec{y}],\phi),
\end{equation}
for every $\phi\/$. In this case the problem simplifies considerably
because once $\esp[\ranvec{y}]\/$ and $\esp[D(\ranvec{y},\bm{\mu}^*)]\/$ have been found (most
likely by simulations), then it is possible to use standard methods
to find $\phi^*\/$, since \eqref{eq:equivalence-minimize-phi} becomes
\begin{equation}
  \label{eq:minentropy-specialcase}
  \riskf_2(\phi)=H(\esp[\ranvec{y}],\phi) + \frac{1}{2\phi}\esp[D(\ranvec{y},\bm{\mu}^*)]\/,
\end{equation}
which is simple to evaluate.

A case worth mentioning when \eqref{eq:simplified-phi-star} occurs is
when the response distribution is a proper dispersion model (see
\citet[Chap. 5]{Jorgensen-book}), i.e.~when $c$ in \eqref{eq:mean-value-par} can be decomposed as
\begin{equation}
  \label{eq:proper-dispersion}
  c(y,\phi) = d(y)e(\phi)\/,
\end{equation}
for some functions $d\/$ and $e\/$. Then, the first term on the right hand
side of \eqref{eq:equivalence-minimize-phi} becomes
\begin{align*}
  -\esp[\log(C(\ranvec{y},\phi))] &= -\esp\left[\prod_{i=1}^m \log\left(c \left(Y_i,\frac{\phi}{w_i}\right) \right)\right] \\
                              &= -\esp\left[\prod_{i=1}^m \log\left(d(Y_i) e\left(\frac{\phi}{w_i}\right) \right)\right] \\
                              &= - \sum_{i=1}^m \esp[\log(d(Y_i))]
                                - \sum_{i=1}^m \log\left( e \left(\frac{\phi}{w_i}\right)\right)\/.
\end{align*}
Since $\sum_{i=1}^m \esp[\log(d(Y_i))]\/$ does not depend on $\phi\/$, the
problem reduces to minimizing
\begin{equation*}
  \riskf_3(\phi) =
  -\sum_{i=1}^m\log\left(e\left(\frac{\phi}{w_i}\right)\right) +
  \frac{1}{2\phi}\esp[D(\ranvec{y},\bm{\mu}^*)]\/,
\end{equation*}
which can be done using standard optimization methods. Now, it is
known that there are only three exponential dispersion models for
which the factorization in \eqref{eq:proper-dispersion} holds: the
gamma, inverse Gaussian and normal distributions (this result is
commented in \citet[Chap. 5]{Jorgensen-book} and proven in Daniels
\cite{daniels-saddlepoint}). Table \ref{tab:e-pedf} gives $e\/$ is for
these three models.

\begin{table}[h]
  \centering
  $\begin{array}{|c|c|c|c|}
     \hline
     \mbox{\bf Distribution} & \mbox{Normal} & \mbox{Gamma} & \mbox{Inverse Gaussian} \\
     \hline
     e(\phi) & \phi^{-1/2} &\displaystyle
                             \frac{e^{-1/\phi}}{\Gamma(\frac{1}{\phi})
                             \phi ^{1/\phi}}  &  \phi^{-1/2} \\ 
     \hline
  \end{array}
  $
  \caption{$e(\phi)\/$ for the three proper exponential dispersion
    families}
  \label{tab:e-pedf}
\end{table}

Let us now consider the general case where
\eqref{eq:minentropy-specialcase} does not hold. Again MCMC methods can
be helpful. Let $\ranvec{y}^1,\ldots,\ranvec{y}^N\/$ be $N\/$ simulations 
of $\ranvec{y}\/$ from the posterior predictive distribution 
(superscripts are used since $Y_i\/$ was already defined to be the 
$i$-th entry of $\ranvec{y}\/$). Now define
\begin{equation*}
  \tilde{\riskf}_N(\phi) = - \frac{1}{N}\sum_{i=1}^N \log(C(\ranvec{y}^i,\phi))
  + \frac{1}{2 \phi N} \sum_{i=1}^N D(Y^i,\bm{\mu}^*)\/,
\end{equation*}
then, for every fixed $\phi\/$
\begin{equation}
  \label{eq:approx-r2}
  \lim_{N\rightarrow \infty} \tilde{\riskf}_N(\phi) =
  \riskf_2(\phi)\qquad \mbox{a.s.}\/.
\end{equation}
Let $\tilde{\phi}_N=\argmin \tilde{\riskf}_N (\phi)\/$. Since
$\tilde{\riskf}_N\/$ is simple to evaluate with a computer, standard
univariate optimization methods can be used to find
$\tilde{\phi}_N\/$. The question now is whether $\tilde{\phi}_N\/$ converges
to $\phi^*\/$ as $N\rightarrow \infty\/$? We have not found easy--to--check 
sufficient conditions that guarantee convergence,
although the following theorem might be useful in some cases.

\begin{proposition}
  If the convergence in \eqref{eq:approx-r2} is uniform almost surely
  w.r.t. $\phi$, then
  \begin{equation*}
    \riskf_2(\phi^*) = \lim_{N\rightarrow\infty} \tilde{\riskf}_N(\tilde{\phi}_N)
	\qquad a.s.
  \end{equation*}
\end{proposition}
\begin{proof}
On the one hand we have that for  every $n\in\nat\/$
  \begin{align*}
    \riskf_2(\phi^*) &\leq \riskf_2(\tilde{\phi}_n)\\
    \therefore\qquad \riskf_2(\phi^*) &\leq  \liminf_n \riskf_2(\tilde{\phi}_n)\/.\\
  \end{align*}
  On the other hand, let $\epsilon>0\/$, since
  $\tilde{\riskf}_N\rightarrow \riskf_2\/$ uniformly a.s., then with
  probability one there exists $M>0\/$, such that for every $n\geq
  M\/$,
\begin{equation*}
  |\tilde{\riskf}_n(\tilde{\phi}_n) - \riskf(\tilde{\phi}_n) | < \epsilon\/.
\end{equation*}
By the definition of $\tilde{\phi}_n\/$,
\begin{equation*}
  \tilde{\riskf}_n(\tilde{\phi}_n) \leq \riskf_n(\phi^*)\/,\qquad
  \mbox{for all } n\in\nat\/.
\end{equation*}
Then for every $n\geq N\/$, 
\begin{equation*}
  \riskf_2(\tilde{\phi}_n) - \epsilon < \riskf_n(\phi^*)\/,
\end{equation*}
thus
\begin{align*}
  \limsup_n \riskf_2(\tilde{\phi}_n) - \epsilon & \leq \limsup \riskf_n(\phi^*)\\
  \therefore \qquad \limsup_n \riskf_2 (\tilde{\phi}_n) - \epsilon & \leq \riskf_2(\phi^*)\qquad a.s.
\end{align*}
Since this is true for $\epsilon>0\/$, this implies that
\begin{equation*}
  \limsup_n \riskf_2(\tilde{\phi}_n) \leq \riskf_2(\phi^*)\/,
\end{equation*}
and therefore
\begin{equation*}
  \riskf_2(\phi^*) = \lim_{n\rightarrow\infty}\riskf(\tilde{\phi}_n) \qquad a.s.
\end{equation*}
\end{proof}

\section{On the Applicability of the Entropic Premium}

The previous section showed how one can find the entropic premium 
of a GLM, theoretically. In this section we address its
practicability. In other words, we address the following question: is
entropic credibility feasible for real--life datasets?

From Proposition \ref{prop:entropic-beta} and Corollary
\ref{cor:entropic-mean}, we have that once the response distribution,
explanatory variables and prior have been chosen, the following steps
give the entropic premium:

\begin{enumerate}
\item Find $\esp[\ranvec{y}]\/$ (see the paragraph preceding Proposition
  \ref{prop:entropic-beta} for the definition of $\esp[\ranvec{y}]\/$).

\item Fit a frequentist GLM with the same covariates, response
  distribution and weights, but with observed response vector equal to
  $\esp[\ranvec{y}]\/$. This gives $\bm{\beta}^*\/$, the entropic estimator of the
  coefficients.

\item Find the entropic mean using \eqref{eq:entropic-premium}.
\end{enumerate}

Steps 2 and 3 are simple: one can perform these computations in
R (see \cite{R}) without major problems. The difficult part is Step 1. To the
best of our knowledge, the simplest way to solve this problem is using
Markov Chain Monte Carlo (MCMC). In the paragraphs that follow we give
some recommendations on how to use this method.

It is important to consider that the greater $m\/$ (the dimension of
$\esp[\ranvec{y}]$), the more demanding the computations (both in terms of
memory and CPU). Thus, it is very useful to first aggregate the data as in
\eqref{eq:aggregate-equation}. This can drastically reduce $m\/$ and
turn an infeasible computation into something manageable.

A continuous variable can make data aggregation useless, especially
when this happens with several different variables in the
dataset. In such cases one should consider converting the support of 
these variables into intervals, hence transforming them into categorical 
variables.

Using Bayesian methods for variable selection can be time
consuming. This is because one would need to run MCMC simulations for
each combination of variables. A pragmatic approach to deal with
this is to choose the variables using a frequentist GLM (which is
much faster to fit). The resulting combination of variables can be
used to build a starting model in the Bayesian case.

\subsection{Illustrative Example}

In this section we use the R interface to STAN
(see \cite{rstan}) to find an entropic credibility estimate of a
severity model for a publicly available dataset. The main purpose of
this example is to show that it is feasible to obtain entropic credibility
premiums. We leave out the discussion about the convergence of the MCMC and
the goodness of fit of the model. The interested reader can find the 
commented R code used for this example at
\url{https://gitlab.com/oquijano/bayesian-credibility-glms}.

{\linespread{1}
\small
\begin{table}[h]
  \centering
  \begin{tabular}[h]{|ll|}
    \hline
    Variable name & Description\\
    \hline
    \hline
    \verb+veh_value+&	vehicle value, in \$10,000s \\
    \hline
    \verb+exposure+ & 0-1 \\
    \hline
    \verb+clm+ & occurrence of claim (0 = no, 1 = yes)\\
    \hline
    \verb+numclaims+ & number of claims\\
    \hline
    \verb+claimcst0+ & 	claim amount (0 if no claim)\\
    \hline
    \verb+veh_body+ & vehicle body, coded as\\				
                    & BUS\\
                    & CONVT = convertible  \\
                    & COUPE   \\
                    & HBACK = hatchback                  \\
                    & HDTOP = hardtop\\
                    & MCARA = motorized caravan\\
                    & MIBUS = minibus\\
                    & PANVN = panel van\\
                    & RDSTR = roadster\\
                    & SEDAN    \\
                    & STNWG = station wagon\\
                    & TRUCK           \\
                    & UTE - utility\\
    \hline
    \verb+veh_age+ & age of vehicle: 1 (youngest), 2, 3, 4\\
    \hline
    \verb+gender+ & gender of driver: M, F \\
    \hline
    \verb+area+ & driver's area of residence: A, B, C, D, E, F \\
    \hline
    \verb+agecat+ & driver's age category: 1 (youngest), 2, 3, 4, 5, 6\\
    \hline
  \end{tabular}
  \caption{Vehicle insurance variables}
  \label{tab:vehicle-insurance-description}
\end{table} }

The dataset appears in de Jong and Heller \cite{glm-insurance-book}. It is
based on 67,856 one--year policies from 2004 or 2005. It can be downloaded 
from the companion site of the book:
\url{http://www.acst.mq.edu.au/GLMsforInsuranceData}, as the
dataset called Car. Table \ref{tab:vehicle-insurance-description}
shows the description of the variables as provided at the website.

We use a GLM with gamma responses to model the severity
(variable \verb+claimcst0+). We modified two explanatory variables, dividing 
the support of the continuous variable \verb+veh_value+ into three intervals
$[0,1.2)\/$,$[1.2,1.86)\/$ and $[1.86,\infty)\/$, which we label as
\verb+P1+, \verb+P2+ and \verb+P3+, respectively. The areas of residence 
A,B,C and D were also grouped together, thus the variable \verb+area+ 
included in our model takes three values: ABCD, E or F.

\begin{table}[h]
  \centering
  \begin{tabular}{|rp{3cm}|rl|}
    \hline
    \multicolumn{2}{|c|}{Model information}& \multicolumn{2}{c|}{MCMC
                                           information}  \\
    \hline
    \bf{Response distribution} & gamma & \bf{No. of chains} & 3\\
    \bf{Weight variable} & \verb+numclaims+ & \bf{Warmup period} &  $2,000$ \\
    \bf{Covariates} & \verb+agecat+(\verb+1+) & & \\
    & \verb+gender+(\verb+F+) &  \bf{Simulations kept} & $28,000$  \\
    & \verb+area+(\verb+ABCD+) & \bf{(per chain)} & \\
    & \verb+veh_value+(\verb+P1+) & & \\
    \bf{Prior} & betas are i.i.d. $U(-20,20)$ and $\phi\sim U(0,1000)$. & & \\
    \hline
  \end{tabular} 
  \caption{Severity model.}
  \label{scheme-severity-model}
\end{table}

There is no out--of--sample information for this example and
therefore we use non--informative priors for all the parameters (see Remark \ref{rem:prior}): The
beta regression coefficients are assumed to be i.i.d.~and to follow 
a uniform distribution on the interval $(-20,20)\/$. The dispersion 
parameter is assumed to follow a uniform distribution on $(0,1000)\/$, 
independently from the betas.

After aggregating the data (using \eqref{eq:aggregate-equation}), the
number of observations are reduced from $67,856\/$ policies to $101\/$ 
classes. Table \ref{scheme-severity-model} shows the information used 
for the Bayesian severity GLM. The value between parenthesis on the
right of each explanatory variable corresponds to the reference
category used in the model.

As shown in Proposition \ref{prop:entropic-beta}, in order to find the
entropic estimator for the betas, it is first necessary to find the
posterior mean of $\esp[\ranvec{y}]$\/. This is where MCMC is
needed. For this example the simulations on STAN took around fifty
seconds, counting compilation time, on three 2.67GHz processors.
After, the entropic betas are found by fitting a frequentist GLM with
the posterior mean as response vector. Table \ref{tab:entropic-betas}
shows the entropic coefficients obtained for this example.


\begin{table}[h]
\centering
\begin{tabular}{llll}
  \hline
 Coefficient& Value  &  Coefficient& Value  \\ 
  \hline
(Intercept) & 7.784 & genderM & 0.183 \\ 
  agecat2 & -0.207 & areaE & 0.152 \\ 
  agecat3 & -0.303 & areaF & 0.377 \\ 
  agecat4 & -0.301 & veh\_valueP2 & -0.117 \\ 
  agecat5 & -0.403 & veh\_valueP3 & -0.156 \\ 
  agecat6 & -0.331 &  &  \\ 
   \hline
\end{tabular}
\caption{Entropic coefficients}
\label{tab:entropic-betas}
\end{table}

\begin{figure}[h]
  \centering
  \includegraphics[scale=0.5]{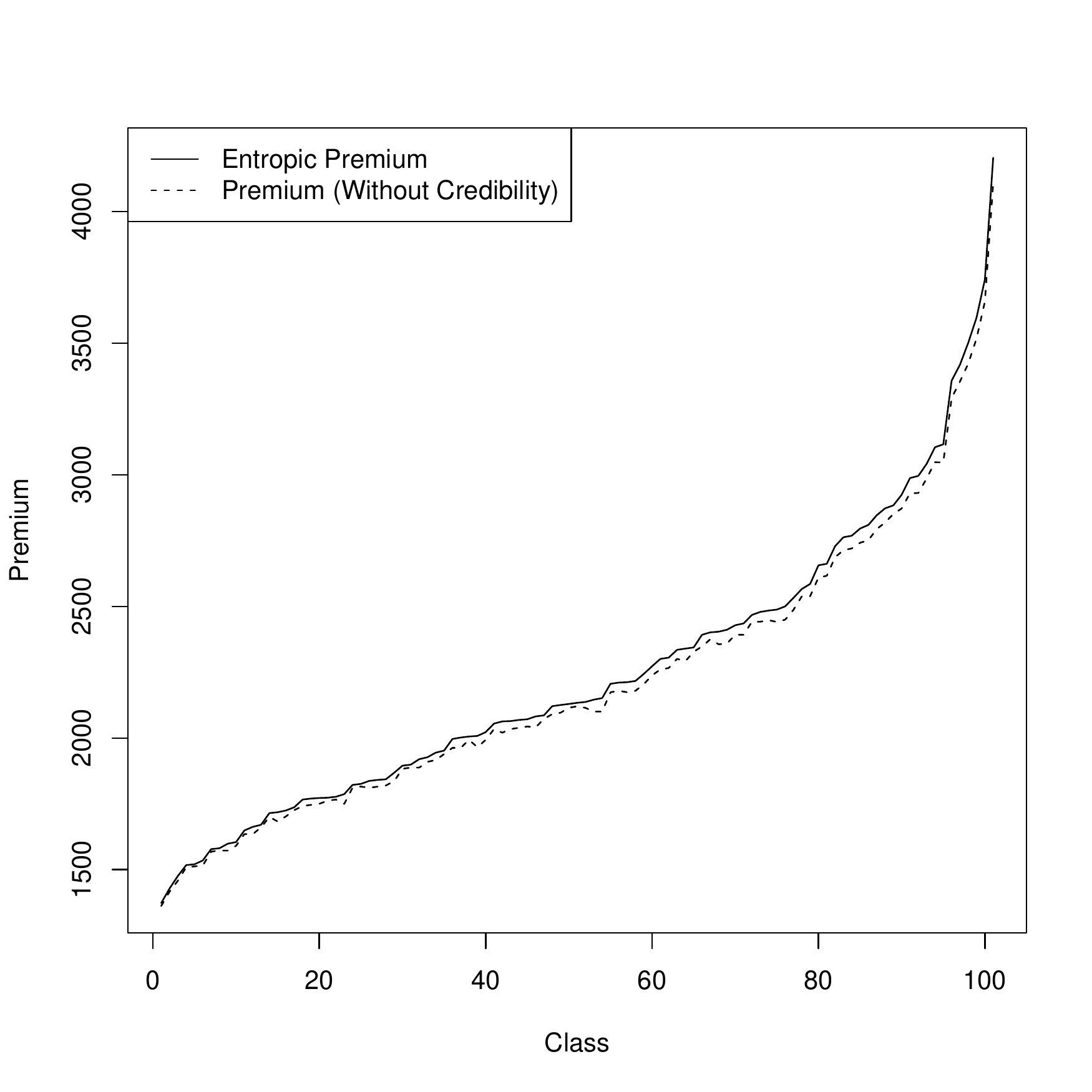}
  \caption{Entropic credibility premiums in increasing order}
  \label{fig:example-entropic-premiums}
\end{figure}

Corollary \ref{cor:entropic-mean} can be used to obtain the entropic
premium for each homogeneous class. We do not show here a full table
of premium values since there are 101 classes and it would take too
much space. Nevertheless, a full table can be found at this article's
website in the section ``Classes Table''. Figure
\ref{fig:example-entropic-premiums} shows a graph of the entropic
premiums in increasing order for all the classes in this example and
compares it to premiums obtained using a frequentist GLM without any
credibility considerations. We see that for some risk classes the GLM 
premiums are fully credible while for others the entropic credibility 
premiums are larger and more conservative.

	\begin{remark}
          The use of the uniform prior here is arbitrary and for
          illustrative purposes only. The goal here is not to seek the
          best Bayesian analysis to this particular dataset, but
          rather to illustrate the feasibility of our method. That is,
          to show that it can be used with real--life datasets and get the
          results in a reasonable amount of time. It is up to the user
          of the method to choose a prior based on the criterion of
          their preference.
	\label{rem:prior}
	\end{remark}

\section*{Conclusion}
As a Bayesian model, linear credibility (described at the beginning of
the introduction) is rather artificial: the adequacy of Jewell's prior
in any given situation is never discussed and the main focus is to
ensure a credibility premium that is easy to compute. This
convenience was crucial when B\"uhlman and Jewell originally published
their work since computing power was scarce and expensive. Nowadays,
not only computing power is cheap, but also sophisticated simulation
software is available to anyone on the internet.

We propose then a modern Bayesian approach to credibility. In this
way one can choose a prior based on out--of--sample information rather
than on ease of computation. The limitation of possible priors is now
set by the convergence of MCMC simulations. We use the relative
entropy between the ``true'' model and the estimated one as the loss
function.

Our proposal, when compared to classical credibility results for the
exponential families, has the additional advantage of considering the 
uncertainty on the dispersion parameter. Finally, applying our method to a
publicly available dataset shows that although substantial
computations are required to obtain the credibility estimates, it
is possible to apply this method to real--life datasets. 

\section*{Acknowledgments}

The authors are sincerely grateful to the anonymous reviewers for their 
constructive comments. We dedicate this paper to the memory of Bent
J{\o}rgensen whose contribution to the theory of exponential dispersion 
families opened the way for future developments. 

\clearpage
\bibliographystyle{plainnat}
\bibliography{referencias}

\appendix

\section*{Appendix: Proofs of technical lemmas}

{\it Proof of Lemma \ref{lem:unit-deviance-decomposition}}

\noindent
Let $d\/$ be the unit deviance function of the response distribution
and $y,\mu\in\Omega\/$. By \eqref{eq:regular-unit-deviance}, we have
that
\begin{align*}
  d(y,\mu)& = 2\left[ y \left\{ \dot{\kappa}^{-1}(y) -
            \dot{\kappa}^{-1}(\mu) \right\} - 
            \kappa(\dot{\kappa}^{-1}(y)) +
            \kappa(\dot{\kappa}^{-1}(\mu))
            \right]\\
          & = 2\left[ y \dot{\kappa}^{-1}(y)-
            \kappa(\dot{\kappa}^{-1}(y)) \right] +
            2\left[\kappa(\dot{\kappa}^{-1}(\mu)) - 
            y \dot{\kappa}^{-1}(\mu)   \right]\\
          & =d_1(y) + d_2(y,\mu)\/,
\end{align*}
where $d_1(y)= 2\left[ y \dot{\kappa}^{-1}(y)-
  \kappa(\dot{\kappa}^{-1}(y)) \right]\/$ and
$d_2(y,\mu) =2\left[\kappa(\dot{\kappa}^{-1}(\mu)) - y
  \dot{\kappa}^{-1}(\mu)\right]\/$. Now, let $Y\/$ be a random variable
with support in $\Omega\/$ and $\mu\in\Omega\/$ be fixed, then
\begin{align*}
  \esp[d_2(Y,\mu)] & = \esp\left[ 2\left[\kappa(\dot{\kappa}^{-1}(\mu)) - Y
                     \dot{\kappa}^{-1}(\mu)\right] \right]\\
                   & = 2\left[\kappa(\dot{\kappa}^{-1}(\mu)) - \esp[Y]
                     \dot{\kappa}^{-1}(\mu)\right]\\
                   & = d_2(\esp[Y],\mu)\/.
\end{align*}
\hfill$\Box$\\

\noindent
{\it Proof of Lemma \ref{lem:d_2-minimization}}

\noindent
  The first part is a direct consequence of
  \eqref{eq:unit-deviance-decomposition}. Since $y\/$ is fixed,
  minimizing the right hand side is equivalent to minimizing $d_2\/$ and
  therefore the claim is true.

  The unit deviance $d\/$ is such that (see Chapter 1 of J{\o}rgensen 
  \cite{Jorgensen-book}) $d(y,\mu)>0\/$ if $y\neq \mu\/$ and $d(y,\mu)=0\/$
  when $y=\mu\/$. Thus $d(y,\mu)\/$ is minimized when $y=\mu\/$ and by Part
  1 above the same applies to $d_2(y,\mu)\/$.

\hfill$\Box$

\noindent
{\it Proof of Lemma \ref{lem:glm-deviance-decomposition}}

\noindent
  From the definition of $D\/$ and Lemma
  \ref{lem:unit-deviance-decomposition}, we have that
    \begin{align*}
    D(\bm{y},\bm{\mu})&=\sum_{i=1}^mw_id(y_i,\mu_i)\\
    &=\sum_{i=1}^m w_i (d_1(y_i)+d_2(y_i,\mu_i))\\
    &=\sum_{i=1}^mw_i d_1(y_i) + \sum_{i=1}^m w_i d_2(y_i,\mu_i)\\
    &=D_1(\bm{y}) + D_2(\bm{y},\bm{\mu})\/,
  \end{align*}
  where $D_1(\bm{y})=\sum_{i=1}^m w_i d_1(y_i)\/$ and
  $D_2(\bm{y},\bm{\mu})=\sum_{i=1}^m w_id_2(y_i,\mu_i)\/$. This proves
  \eqref{eq:deviance-decomposition}. Let $\ranvec{y}=(Y_1,\ldots,Y_m)\/$ be a
  random vector with support on $\Omega^m\/$, and $\bm{\mu}\in\Omega^m\/$ be
  fixed. Then
  \begin{equation*}
    \esp[D_2(\ranvec{y},\bm{\mu})]=\esp[ \sum_{i=1}^m w_i d_2(Y_i,\mu_i)  ]
    =\sum_{i=1}^m w_i d_2(\esp[Y_i],\mu_i) =D_2(\esp[\ranvec{y}],\bm{\mu}),
  \end{equation*}
  which proves \eqref{eq:d2-property}.
\hfill$\Box$

\end{document}